\documentclass[aps,prx,twocolumn,groupedaddress,showpacs,superscriptaddress,scrartcl,longbibliography]{revtex4-2}
\usepackage{xcolor,lineno,mathtools,comment,mathrsfs,natbib,amssymb,amsmath,graphicx,epstopdf,hyperref,amsthm,dsfont,enumerate,bm}

\newtheorem*{lemma6*}{Lemma 6}
\newtheorem*{theorem1*}{Theorem 1}

\newtheorem*{lemmaaA*}{Lemma A}
\newtheorem*{lemmaaB*}{Lemma B}

\newtheorem{condition}{Condition}

\newtheorem*{postulate1'*}{Postulate 1'}
\newtheorem*{postulate1*}{Postulate 1}
\newtheorem*{postulate3'*}{Postulate 3'}

\newcommand{\damian}[1]{{\color{black}{#1}}}	
\newcommand{\damiann}[1]{{\color{black}{#1}}}			
\newcommand{\adrian}[1]{{\color{black}{#1}}}			

\newcommand{\adriann}[1]{{\color{black}{#1}}}			
\newcommand{\damiannn}[1]{{\color{black}{#1}}}	

\newcommand{\dpg}[1]{{\color{black}{#1}}}
\newcommand{\dpgg}[1]{{\color{black}{#1}}}

\begin{document}
\interfootnotelinepenalty=10000

\title{\adriann{Testing} the nonclassicality of spacetime: what can we learn from \dpgg{Bell--Bose \emph{et al}.-Marletto-Vedral} experiments?}

\author{Adrian Kent}
\email[]{A.P.A.Kent@damtp.cam.ac.uk}
\affiliation{Centre for Quantum Information and Foundations, DAMTP,
Centre for Mathematical Sciences, University of Cambridge,
Wilberforce Road, Cambridge, CB3 0WA, United Kingdom}
\affiliation{Perimeter Institute for Theoretical Physics, 31 Caroline Street North, Waterloo, ON N2L 2Y5, Canada}

\author{Dami\'an Pital\'ua-Garc\'ia}
\email[]{D.Pitalua-Garcia@damtp.cam.ac.uk}
\affiliation{Centre for Quantum Information and Foundations, DAMTP,
Centre for Mathematical Sciences, University of Cambridge,
Wilberforce Road, Cambridge, CB3 0WA, United Kingdom}

\date{\today}


\begin{abstract}

The \dpgg{Bose \emph{et al}.-Marletto-Vedral} (BMV) experiment [S. Bose \emph{et al}., Phys. Rev. Lett. 119, 240401 (2017); C. Marletto and V. Vedral, Phys. Rev. Lett. 119, 240402 (2017)]
aims to prove that spacetime is nonclassical by observing entanglement generated by gravity. However, local hidden variable theories (LHVTs) can simulate the
entangled correlations. We propose to extend the entanglement generated by the BMV experiment to distant quantum particles in a Bell experiment. Violating a
Bell inequality would rule out LHVTs, providing a stronger proof of the nonclassicality of spacetime than the BMV proposal.

\end{abstract}

\maketitle

\section{Introduction}
We do not yet know how to unify our two most fundamental theories \damian{of physics}, quantum theory (QT) and general relativity (GR).
  Each is supported by overwhelming experimental evidence in its respective domain.
  However, they are formulated very differently.   GR is an essentially classical
  theory in which the stress-energy tensor\damiann{, the spacetime geometry and all} physical quantities take well-defined values;
  QT predicts that distinct mass-energy configurations can be in quantum superposition.
  One popular approach to resolving this tension is to seek a quantum theory of gravity\damian{,}
  such as string theory \cite{GreenSchwarzWittenbook} or loop quantum gravity
  \cite{Rovellibook}.   It has even been claimed that there is no consistent
  alternative 
  \damian{(e.g., \cite{DR11,EH77,T06,MV17.2})}. However, this has also often been questioned (e.g. \cite{C08}).
  In particular, \damiann{the argument of Ref. \cite{EH77} has} been refuted (e.g., \cite{HC01,M06,AKR08,K18}). 
   It seems the question can only be settled empirically.

In their simplest form, semi-classical gravity models \cite{M62,R63,K78,KR80} 
are defined by taking
\begin{equation}\label{scg}
G_{\mu \nu} = \dpg{\kappa}\langle \hat{T}_{\mu \nu} \rangle  \, , 
\end{equation}
where the left hand side is the Einstein tensor of classical \damian{spacetime}, and the right hand side is
the expectation value of the stress-energy tensor of quantum matter propagating in that \damian{spacetime}, \dpg{with $\kappa$ being a proportionality constant}.
If this is taken to be defined by Everettian quantum theory with purely unitary
evolution, \damian{Eq.} (\ref{scg}) is inconsistent with observation (e.g. \cite{PG81});
it is also unclear that it defines a consistent theory.
However\damian{,} it remains possible that (\ref{scg}) holds in some regime (e.g. \dpg{\cite{CST19,TD16,BW20,K21}}).
For example, given a version of quantum theory with explicit localized collapses,
a classical gravitational field could couple to the local quantum state, which is defined
by the initial conditions, evolution, and collapse events (only) in the \damian{causal past} of the relevant point \cite{K18,K21,
K21.2}. Another possibility is that quantum superpositions of sufficiently distinct
energy-mass configurations are dynamically suppressed \damian{(e.g., \cite{K66,D84,D87,D89,P96,P98,P14,HPF19})}.

All these options are problematic. For example, a recent experiment \cite{DPCDLB21}
gives strong evidence against the \damian{Di\'osi-Penrose proposals \cite{D87,D89,P96,P14,HPF19}} for gravitationally induced collapse and suggests that a radically new approach may be needed to pursue this idea.
However, quantum gravity theories also have well known problems (see e.g. \cite{WR21}
for some discussion).

Recent proposals for table-top tests of quantum gravity,
beginning with so-called \dpgg{Bose \emph{et al}.-Marletto-Vedral} (BMV) experiments
\cite{BMMUTPGBKM17,MV17}, aim to give strong evidence as to whether gravity is mediated by quantum information exchange.
  This would also give indirect evidence that spacetime is quantum \footnote{There are other possibilities.
  For example, one can imagine theories in which gravity is mediated by quantum particles
  but spacetime superpositions are precluded or in which the fundamental theory of spacetime is not described by a quantum formalism.}. The essential idea of BMV's proposal is to place two adjacent mesoscopic masses in position superposition states and allow them to fall along paths
such that one pair of paths is significantly closer than the others,
before recombining the paths interferometrically.    
With appropriate masses, separations, and fall times, which it is hoped
will be experimentally feasible in the foreseeable future, the 
Schr\"odinger evolution with a Newtonian potential implies that
an entangled final state can be generated from a separable initial state.   
As the position degrees of freedom are correlated with internal degrees of freedom,
this entanglement can be tested by measuring entanglement witnesses.

Refinements to the experiments have since been proposed \cite{KMBM20},
as have alternative experiments aiming to witness (non-)quantum behaviour
of gravity by other methods \damiannn{(e.g., \cite{H21,HPF19,HVNCRI21,CMT21})}. 
There is also ongoing debate (e.g. \cite{HR18,MV20,HVNCRI21}) over how definitively the quantum nature of gravity would be demonstrated by the generation of entanglement in a BMV experiment.

A separate issue, our focus here, is how definitively passing the type of entanglement witness
test proposed by BMV would establish that entanglement had indeed been generated.
There is clearly a logical loophole in this inference, since
we know that local hidden variable theories can simulate 
arbitrary entangled correlations in any
experiment -- such as the BMV proposals -- that does not involve
measurements in spacelike separated wings.  
Of course, Bell experiments give very strong -- even if not yet
loophole-free \cite{K05,K20} -- evidence against
local hidden variable theory explanations for correlations
between measurements on entangled matter.
However, Bell experiments to date have not tested
states in which (according to quantum gravity intuitions) entanglement
is generated by gravitational interactions.
As already noted, there is some motivation for considering
models in which classical degrees of freedom associated with gravity
couple to, and might indeed contain complete information about, the local quantum
state.   In principle, this would allow classical simulation of the correlations
that a quantum analysis would ascribe to gravitationally generated entanglement.

We are not aware of any convincing model that would reproduce the correct
correlations for all the entangled states that could (on a quantum analysis)
be generated by varying the experimental parameters.
Still, BMV experiments aim to establish a fundamental
feature of nature, and in our view it is worth striving to eliminate
loopholes in their interpretation, for reasons similar to those
motivating the ongoing quest to remove Bell experiment loopholes. 
Although it is a theoretically familiar idea that the space and time
we inhabit somehow emerge from a more fundamental quantum description,
it remains an extraordinary claim, which justifies extraordinary
care in assessing evidence.

We propose here \adriann{extending} the entanglement generated in a BMV experiment to distant quantum particles in a Bell experiment.
For the reasons just given, this would provide a more definitive test of the nonclassicality of gravity \damiann{and spacetime}. 

\section{A Bell--Bose \emph{et al}.-Marletto-Vedral experiment}

\begin{figure}
\includegraphics[scale=0.286]{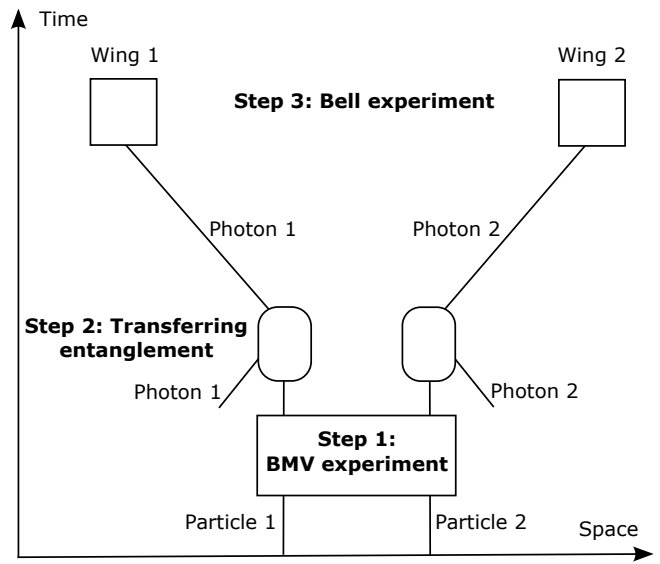}
\caption{\label{fig} \textbf{Our proposed BBMV experiment.} We illustrate our proposed BBMV experiment in a spacetime diagram in $1+1$ dimensions in a background Minkowski spacetime. Our proposed experiment comprises three broad steps. In step 1 (large rectangle), the BMV experiment \cite{BMMUTPGBKM17,MV17} is applied on Particles 1 and 2 (vertical lines). The entanglement generated via the BMV effect is transferred to Photons 1 and 2 (diagonal lines) in step 2 (rounded rectangles). In step 3 (small squares), a Bell experiment \cite{Bell} is applied on Photons 1 and 2, testing (for example) CHSH inequalities \cite{CHSH69}. If conditions \ref{asu1}--\ref{asu4} hold and a Bell inequality is violated then we can exclude explanations via local hidden variables, obtaining a stronger proof for the nonclassicality of the gravitational field than the original BMV experiment. The illustration is not at scale. In particular, we expect the distance between the wings in the Bell experiment to be much larger than the distance between the particles in the BMV experiment.}
\end{figure}

We propose an experimental test for the nonclassicality of gravity comprising three general steps. Broadly, these steps are the following \damian{(see Fig. \ref{fig})}.

\begin{enumerate}
\item Perform the BMV experiment \cite{BMMUTPGBKM17,MV17} with a pair
  of mesoscopic masses with spin. Let us call these systems ``particle
  1" and ``particle 2", and their spin degrees of freedom ``$S_1$" and
  ``$S_2$". Arrange the experiment so that the particles' states are
  initially unentangled, and the only non-negligible interaction
  between the particles during the experiment is the gravitational
  field. The hypothesis that gravity is quantum implies that $S_1$ and
  $S_2$ become entangled due to the gravitational interaction of the
  particles' masses \cite{BMMUTPGBKM17,MV17,CR19}. Assuming this
  hypothesis is correct, arrange the experiment so that the final
  state between $S_1$ and $S_2$ is approximately the
  maximally entangled singlet state:
\damiannn{
\begin{equation}
\label{singlet}
\rho_{S_1S_2}^{(1)}\approx \bigl( \lvert \Psi^{-}\rangle\langle \Psi^{-}\rvert \bigr)_ {S_1S_2},
\end{equation}
where  
\begin{equation}
\label{defsinglet}
\lvert \Psi^{-}\rangle= \frac{1}{\sqrt{2}}\bigl(\lvert0\rangle\lvert 1\rangle-\lvert1\rangle\lvert 0\rangle\bigr).
\end{equation}
}

\item Transfer the quantum state from the spin degrees of freedom
  $S_i$ of particle $i$ to some degrees of freedom $P_i$ (e.g.,
  polarization) of a photon, which we call ``photon $i$", for
  $i=1,2$. Arrange the experiment so that the interaction between
  particle 1 (respectively 2) and photon 1 (2) is local, i.e., without interacting
  with particle 2 (1) or photon 2 (1). If at the end of step 1, the
  joint quantum state of the particles' spins $S_1$ and $S_2$ 
\damiannn{satisfies (\ref{singlet})}, then at the end of this step
  the joint quantum state of the photons' degrees of freedom $P_1$ and
  $P_2$ is 
\damiannn{
\begin{equation}
\label{singlet2}
\rho_{P_1P_2}^{(2)}\approx \bigl( \lvert \Psi^{-}\rangle\langle \Psi^{-}\rvert \bigr)_ {P_1P_2}.
\end{equation}
}

\item Send photons 1 and 2 to distant wings and implement \damian{a Bell \cite{Bell} experiment} on systems $P_1$ and $P_2$ at spacelike separation,
    testing (for example) Clauser-Horne-Shimony-Holt (CHSH) \cite{CHSH69} inequalities.

\end{enumerate}

We call this a Bell--\dpgg{Bose \emph{et al}.-Marletto-Vedral} (BBMV) experiment.
We emphasize the conditions that need to be satisfied experimentally.

\begin{condition}
\label{asu1}
At the beginning of step 1, the states of particles 1 and 2 are initially unentangled.
\end{condition}

\begin{condition}
\label{asu2}
During step 1, the only non-negligible interaction between particles 1 and 2 is gravity.
\end{condition}

\begin{condition}
\label{asu3}
During step 2, transferring the quantum state from the spin $S_i$ of particle $i$ to the degrees of freedom $P_i$ of photon $i$ does not include any interaction between the joint system of particle $i$ and photon $i$ with the joint system of particle $\bar{i}$ and photon $\bar{i}$, either directly or indirectly via another system, for $i=1,2$.
\end{condition}

\begin{condition}
\label{asu4}
The measurements in wing $i$ in the Bell experiment of step 3 are applied on the \damian{degrees of freedom} $P_i$ of photon $i$, for $i=1,2$.
\end{condition}


\damian{Suppose that} the experiment is repeated $N>>1$ times and that conditions \ref{asu1}-\ref{asu4} hold in each run of
the experiment. In this case, observing correlations approximately consistent with CHSH measurements on a quantum singlet,
and so violating the CHSH inequality, in our BBMV experiment would
imply (modulo any remaining loopholes) that the correlations obtained cannot be explained
by locally causal hidden variable models \damian{\cite{Bell,CHSH69}.   We see relatively little motivation for considering non-locally causal hidden
variable models in the context of unifying quantum theory and general relativity, and in particular
for considering such models as an explanation for correlations observed in the
specific BBMV experiment described.    The natural inference would thus be that the correlations
are indeed generated by measurements on an approximate singlet state, and hence that the BMV
sub-experiment did indeed generate entanglement.}

\section{Discussion}

BMV experiments involve testing for correlations that
  would arise from measurements on the entangled quantum states
  that should be generated, according to non-relativistic quantum
  theory using Newtonian gravitational potentials.   
  In principle, such correlations could also be produced by
  a non-quantum theory of gravity involving local hidden
  variables.   If our proposed experiment produced results
  consistent with quantum theory and the BMV analyses, it
  would exclude the latter explanation, and hence provide stronger
  evidence for quantum gravity.   This is not to diminish the
  importance of BMV's \damian{\cite{BMMUTPGBKM17,MV17}} crucial insight.   Arguably, given most
  theorists' Bayesian priors, our proposed experiment would
  further (beyond a BMV experiment) enhance their credence in quantum gravity
  only marginally.   Still, given our incomplete understanding of nature,
  empirical proof is preferable to confident priors.

Ideally, all else being equal, the Bell experiment of step 3 should be as loophole free as possible.
However, we see more motivation for closing some loopholes than others.
There is very strong motivation for closing the locality \damian{loophole \cite{Bell}}, since BBMV experiments are motivated by the
concern that gravitational effects might be mediated by locally causal hidden variables.
Ideally,  the collapse locality loophole \cite{K05,K20} should also
be closed, since one reason for considering this loophole is the possibility that measurement events
occur only when their outcomes leave a record in the gravitational field and hence, in order to
verify genuine non-locally causal correlations, we need to ensure these records are created
in \damian{spacelike} separated regions.
Although this is more challenging, recent advances in experimental quantum information science in space (e.g.,
\dpg{\cite{Yinetal17etal,Liaoetal17etal,SpaceQuest18etal,SJGBLM21etal,KABBBDMOPR21etal,GBCDBKPU21}}) suggest
that the collapse locality loophole could be closed in the foreseeable future
\cite{K20}.

It would also be worthwhile to close the freedom-of-choice loophole \damian{\cite{Bell}}, to eliminate any possibility
that information about choices made at earlier times on one wing propagates via locally causal hidden variables
(which again might be associated with the gravitational field) to influence outcomes on the other.
We see less motivation for closing the detector efficiency loophole \cite{Pearle70};
we find it harder to imagine a plausible theory in which locally causal hidden variables affect
the behaviour of detectors in BBMV experiments but not in standard Bell experiments.

It is hoped that the BMV experiment could be feasible in the foreseeable
future (e.g., \cite{BMMUTPGBKM17,MV17,KMBM20}). Our proposed
BBMV experiment appears not substantially more challenging to implement than the
BMV experiment and a Bell experiment. State of the art
techniques for spin-photon coupling (e.g.,
\cite{MBPZTBP18,SZKBSMBSV18}) suggest that step 2 in our experiment
could be possible in practice.

\damian{
\begin{acknowledgments}
The authors acknowledge financial support from the UK Quantum
Communications Hub grant no. EP/T001011/1. A.K. is partially supported
by Perimeter Institute for Theoretical Physics. Research at Perimeter
Institute is supported by the Government of Canada through Industry
Canada and by the Province of Ontario through the Ministry of Research
and Innovation.  \adrian{A.K. thanks Fay Dowker for helpful comments.  }

  \end{acknowledgments}

}
%

\end{document}